\documentclass[condensedmatter,article,accept,pdftex,moreauthors]{Definitions/mdpi} 
\firstpage{1} 
\pubvolume{1}
\issuenum{1}
\articlenumber{0}
\pubyear{2023}
\copyrightyear{2023}
\externaleditor{{Academic Editor: Amir-Abbas Haghighirad} 
}
\datereceived{4 April 2023} 
\daterevised{28 April 2023}
\dateaccepted{6 May 2023} 
\datepublished{} 
\hreflink{https://doi.org/} 

\Title{Quadriexciton Binding Energy in Electron--Hole Bilayers}
\TitleCitation{Quadriexciton Binding Energy in Electron--Hole Bilayers}

\Author{Cesare Malosso  $^{1}$\orcidA{}, ~Gaetano Senatore $^{2}$\orcidB{} and Stefania De Palo $^{2,3}$\orcidC{}$^*$}

\AuthorNames{Cesare Malosso, Gaetano Senatore and Stefania De Palo}

\AuthorCitation{Malosso, C.; Senatore, G.; Palo, S.D.}
\address{%
$^{1}$ \quad SISSA---Scuola Internazionale Superiore di Studi Avanzati, 34136 Trieste, Italy; 
cmalosso@sissa.it\\
$^{2}$ \quad Dipartimento di Fisica, Universita` di Trieste, Strada Costiera 11, 34151 
Trieste, Italy; senatore@units.it\\
$^{3}$ \quad {CNR-IOM-DEMOCRITOS,} I-34136 Trieste,   Italy; depalo@iom.cnr.it} 

\corres{Correspondence: senatore@units.it, (S.D.P.)}

\abstract{ Excitonic condensation and superfluidity have recently received a renewed attention, due to the fabrication of bilayer systems in 
which electrons and holes are spatially separated and form stable pairs known as indirect excitons. Dichalcogenides- and graphene-based 
bilayers are nowadays built and investigated, giving access to systems with (i) only spin degeneracy and (ii) spin and valley degeneracy. 
Simulation studies performed in the last decades at $T=0$ for simple, model electron--hole bilayers, as function of the interlayer distance and in-layer carrier density, have revealed in case (i) the formation of biexcitons in a tiny region of the parameter space and in case (ii) the 
formation of stable compounds made of four electrons and four holes (quadriexcitons) in a sizable region of the parameter space. Of some interest is the relation of the properties of isolated biexcitons (quadriexcitons) and those of their finite-density counterpart. In fact, the isolated biexciton has been repeatedly studied in the last years with simulations and other techniques. No simulations, instead, are available to our knowledge for the isolated quadriexciton, for which we present here results of the first quantum Monte Carlo (QMC) study. Stability with respect to the dissociation into biexcitons and the pair correlations while varying the interlayer distance $d$ are discussed.}
\begin{document}
\section{Introduction}

An exciton, a bound electron--hole pair in a semiconductor \cite{AM}, is an elementary optical excitation or quasiparticle in solid state 
physics, reminiscent of positronium or hydrogen in atomic physics.  Here we are interested in Mott-Wannier excitons.  Interactions between two excitons can result in the formation of an excitonic molecule \cite{Lampert, Moska}, a new quasi-particle known as biexciton \cite{Moska}. The study of excitons and biexcitons is most naturally performed within the envelope function and effective mass approximations \cite{Ihn}, which provide an effective Hamiltonian for interacting electron and hole quasi-particles, near energy band extrema. Whenever a band has multiple equivalent minima (maxima) the description of electrons (holes) implies a new discrete quantum number, the valley index $v$, which identifies a minimum (maximum). The valley index can be formally treated as a pseudospin; the number $g_v$ of equivalent valleys fixes the length $\tau$ of the pseudospin, $2\tau+1=g_v\rightarrow \tau=(g_v-1)/2$. We can introduce a flavor (or valley-spin) index $\sigma=( \tau_z,s_z)$ to fix the spin and pseudospin states. As electrons and holes have spin 1/2, the total number of flavors is $N_c=2 g_v $ for each particle type, electron or hole. We assume that the Hamiltonian does not contain spin or pseudospin operators. It is evident that if one considers a finite electron--hole system with $N_e\leq N_c$, $N_h\leq N_c$, with $N_e$ ($N_h$) the number of electrons (holes), the ground state wave function $\Phi$ can be exactly factorized in $\Phi=\Psi \zeta$, with  $\Psi$ a nodeless wave function of the particles' Cartesian coordinates and $\zeta$ a wave function of pseudospin and spin coordinates; $\Psi$ will be symmetric for any pair exchange of electrons (holes) coordinates and $\zeta$ will be antisymmetric for any pair exchange of electrons (holes) coordinates. One can of course choose $\Psi$ to also have other symmetries possessed by the Hamiltonian, provided that this does not spoil the exchange symmetry and $\Psi$ remains nodeless.

Long ago it was observed \cite{kittel1972} that when there are $g_v$ equivalent minima in the electron conduction band, up to $N_c=2g_v$ electrons can occupy the same molecular orbital and therefore electron--hole complexes, where up to $N_c$ excitons are possible; the same is true~\cite{kittel1972} when $N_c$ equivalent maxima are present in the valence band. Clearly, in a system with $g_v=2$, such as coupled graphene bilayers \cite{peralietal13,deanGBprl,tutucGBprl,liunpCGB2017, linpCBG2017,tutucprl2018}, $N_c=4$ and quadriexcitons should be possible.  We recall that the problem of an isolated biexciton has already been \mbox{studied~\cite{Tan,Meyertholen_Fogler,Lee}} with DMC and other techniques. 

Before tackling below the treatment of an isolated quadriexciton (or biexciton ) at $T=0$, in the simplest, model electron--hole bilayer (the 
paramagnetic, symmetric bilayer: $m_e^*=m_h^*=m_b$), we briefly summarize here the results of available QMC simulations of systems with finite in-layer carrier density $n$ \cite{depalo2002,maezono2013,quadri}, which we specify via the dimensionless $r_s$ parameter, defined by $\pi r_s^2{a_B^*}^2=1/n$ where $a_B^* =4\pi \epsilon_0\epsilon \hbar^2/(m_b e^2)$. 

The first simulations \cite{depalo2002} were performed in the region of parameter space $0<d/(r_sa_B^*)\leq 3$, $0\leq r_s\leq 30$) with $g_v=1$. It was found that at large density $r_s\lesssim1$, the plasma phase was stable at all distances $d$, due to the screening of the Coulomb attraction by carriers. When increasing $r_s$, an excitonic condensate appeared for $d<d_X(r_s)$ with $d_X(r_s)$ an increasing function of $r_s$. For $d>d_X(r_s)$, the plasma phase was stable up to an intermediate density, $r_s\lesssim 20$, while turning into a Wigner crystal for $20 \lesssim r_s\lesssim 30$. 
More recent simulations~\cite{maezono2013}, performed in a much smaller region of the parameter space ($0<d/a_B^*\leq 4$, $0\leq r_s\leq 8$)  with $g_v=1$,  using a much more flexible wave function and up-to-date computer resources, confirmed semiquantitatively the phase boundary between the excitonic and plasma phases as well as the condensate fractions found in \cite{depalo2002}, while finding for $4 \lesssim r_s\lesssim 8 $ a new, stable biexcitonic phase for $d  \lesssim d_{2X}(r_s) = 0.05a_B^*$. 
Finally, the results of QMC simulations for and electron--hole bilayer with $g_v=2$, $0<d/a_B^*\leq 3.5$ and $0\leq r_s\leq 8$ have just been published~\cite{quadri}. As predicted by \cite{kittel1972}, a quadriexcitonic phase was found and in a sizeable region of the parameter space, compared with the region of stability of the biexcitonic phase found for $g_v=1$. The new phase appeared for $r_s\gtrsim 1$ and $0 \lesssim d/ \lesssim d_{4X}(r_s)$, with $ d_{4X}(r_s)$ an increasing function of $r_s$. At a given $r_s \gtrsim 1$  and for  $ d_{4X}(r_s)\lesssim d \lesssim d_X(r_s)$, there was the excitonic phase, followed by the plasma phase for $ d\gtrsim d_{X}(r_s)$. No trace was found of the biexcitonic~phase.

\section{Hamiltonians and Wave Functions}
Within the envelope-function effective mass approximation \cite{Ihn}, the Hamiltonian of a two-dimensional bilayer reads \vspace{-6pt}
\begin{eqnarray}
 H=&-&\frac{\hbar^2}{2m_e^*}\sum_{v,i}{\nabla^2_{e,v,i}} +\frac{1}{2} \sideset{}{'}\sum_{v,v',i, i'}\frac{e^2}{4\pi \epsilon_0\epsilon|{\bf r}_{e,v,i} -{\bf r}_{e,v',i'}|}  \nonumber\\
&-&\frac{\hbar^2}{2m_h^*}\sum_{v,i}{\nabla^2_{h,v,i}}+\frac{1}{2} \sideset{}{'}\sum_{v,v',i, i'}\frac{e^2}{4\pi \epsilon_0\epsilon|{\bf r}_{h,v,i} -{\bf r}_{h,v',i'}|} \nonumber\\
 &-&\sum_{v,v',i, i'}\frac{e^2} {4\pi \epsilon_0\epsilon\sqrt{|{\bf r}_{e,v,i} -{\bf r}_{h,v',i'}|^2+d^2}}.
 \label{H}
\end{eqnarray}

This~Hamiltonian, apart from the assumption of isotropic masses, is quite general: $v$ and $v'$ run over the valleys and $i$ and $i'$ run over the number of electrons or holes in a given valley. A prime in the sum excludes terms with both $v=v'$ and $i=i'$. The paramagnetic, symmetric bilayer is the one in which $m_h^*=m_e^*=m_b$, all valleys present have the same electron (hole) population and in each valley, there is the same number of spin-up and spin-down electrons (holes). In the rest of this paper, we measure distances using the effective Bohr radius $a_B^* =4\pi \epsilon_0\epsilon \hbar^2/(m_b e^2)$ and energies in effective Rydbergs $Ry^*= \hbar^2/(2 m_b {a_B^*}^2)$, so that the Hamiltonian becomes 
\begin{eqnarray}
{ \cal H}=&-&\sum_{v,i}{\nabla^2_{e,v,i}} + \sideset{}{'}\sum_{v,v',i, i'}\frac{1}{|{\bf r}_{e,v,i} -{\bf r}_{e,v',i'}|} 
-\sum_{v,i}{\nabla^2_{h,v,i}}+ \sideset{}{'}\sum_{v,v',i, i'}\frac{1}{|{\bf r}_{h,v,i} -{\bf r}_{h,v',i'}|} \nonumber\\
 &-&\sum_{v,v',i, i'}\frac{2} {\sqrt{|{\bf r}_{e,v,i} -{\bf r}_{h,v',i'}|^2+d^2}}.
 \label{H1}
\end{eqnarray}

As we argued above for $N_e=N_h=N_c\equiv N$, the ground-state wave function $\Psi^0_{NX}({\bf R}_e,{\bf R}_h)$, with ${\bf R}_e$ (${\bf R}_h$) the electrons (holes) coordinates, (i) must be the solution of 
\begin{equation}
{\cal H }\Psi^0_{NX}({\bf R}_e,{\bf R}_h)={E^0}_{NX}\Psi^0_{NX}({\bf R}_e,{\bf R}_h);
 \label{H2}
\end{equation}
(ii) it must be symmetric for any pair exchange of electrons (holes) coordinates; (iii) in the symmetric bilayer, it must be symmetric for the electron--hole exchange, i.e., $\Psi^0_{NX}({\bf R}_e,{\bf R}_h)= \Psi^0_{NX}({\bf R}_h,{\bf R}_e)$ \cite{Tan}. 

For the biexciton and the quadriexciton, we tackle the problem in Equation~(\ref{H2}) resorting to variational and diffusion Monte Carlo 
\cite{daviddmc,cyrusdmc,rmpfoulkes} (VMC and DMC) as implemented in our own code. For a given $N$, at each $d$, an optimal trial function 
$ \Psi^T_{NX}({\bf R}_e,{\bf R}_h;{\bf c})$ is determined by minimizing the variational energy with respect to a number of optimizable parameters ${\bf c} $ \cite{linearmethod,linearmethod1,sorella-sr}.  We then compute the VMC estimates of the properties of interest using a Monte Carlo integration with $|\Psi^T_{NX}|^2$ as the importance function; the DMC estimates are obtained by employing the optimized $\Psi^T_{NX}$ as the guiding function. 
We now turn to the explicit form of $\Psi^T_{NX}$.

\subsection{Polyexciton Wave Function $\Psi^T_{NX}$}
Here, we restrict to polyexcitons made by $N_c$ electrons and $N_c$ holes. 
Given the symmetry requests on the wave function it is natural to write $ \Psi^T_{NX}({\bf R}_e, {\bf R}_h)$ as a symmetrized Jastrow factor. 
We start from the unsymmetrized form
 \begin{linenomath}
\begin{equation}
J({\bf R}_e, {\bf R}_e)=\exp\left[-(1/2)\sum_{\mu,\mu'}\sideset{}{'}\sum_{i_{\mu},j_{\mu'}}
u_{\mu,\mu'}(|{\bf r}_{i_{\mu}}-{\bf r}_{j_{\mu'}}|)\right],
\label{Ja}
\end{equation}
 \end{linenomath}
 embodying two-body pseudopotentials among all particles. Above the {\it species} index, $\mu=(t,\sigma)$ combines the particle type ($t=e,h$) and the flavor $\sigma=( \tau_z,s_z)$; moreover, the primed sum for $\mu'=\mu$ contains only the terms with $i_{\mu}\ne j_{\mu}$. Evidently 
\begin{linenomath}
   \begin{equation}
 J({\bf R}_e, {\bf R}_e)= J_{ee}({\bf R}_e)J_{hh}({\bf R}_h)J_{eh}({\bf R}_e,{\bf R}_h).
  \end{equation}
 \end{linenomath}
 
 Let us inspect the three terms above. As there is only one electron per flavor, only interflavor terms ($\sigma\ne \sigma'$) survive in
\begin{linenomath}
\begin{equation}
J_{ee}({\bf R}_e)=\exp\left[-(1/2)\sum_{\sigma\ne \sigma'}\sum_{i_{e,\sigma},j_{e,\sigma'}}
u_{e,\sigma;e,\sigma'}(|{\bf r}_{i_{e\sigma}}-{\bf r}_{j_{e,\sigma'}}|)\right],
  \end{equation}
 \end{linenomath}
and in fact, the sum over $i_{e,\sigma}$ and $j_{e,\sigma'}$ in the equation above has just one term; thus,
\begin{linenomath}
\begin{equation}
J_{ee}({\bf R}_e)=\exp\left[-(1/2)\sum_{\sigma\ne \sigma'} 
u_{e,\sigma;e,\sigma'}(|{\bf r}_{i_{e\sigma}}-{\bf r}_{j_{e,\sigma'}}|)\right]\equiv 
\exp\left[\sum_{\sigma< \sigma'} 
\phi(|{\bf r}_{i_{e\sigma}}-{\bf r}_{j_{e,\sigma'}}|)\right],
  \end{equation}
 \end{linenomath}
where we chose the interflavor pseudopotentials as being all equal, $u_{e,\sigma;e,\sigma'}(r)=\phi(r)$. Since we are considering a paramagnetic, symmetric bilayer, it immediately follows that 
\begin{linenomath}
\begin{equation}
J_{hh}({\bf R}_h)=
\exp\left[\sum_{\sigma< \sigma'} 
\phi(|{\bf r}_{i_{h\sigma}}-{\bf r}_{j_{h,\sigma'}}|)\right]. 
 \end{equation}
 \end{linenomath}
 
We note that with the choices made, the product $J_{ee}({\bf R}_e)J_{hh}({\bf R}_h)$ is symmetric under any electron--hole pair exchange as well as under any electrons--holes exchange. 
Let us turn now to $J_{eh}({\bf R}_e,{\bf R}_h)$. We have \vspace{-6pt}
\begin{adjustwidth}{-\extralength}{0cm}
\begin{linenomath}
\begin{eqnarray}
J_{eh}({\bf R}_e,{\bf R}_h)&=&\exp\left[-\sum_{\sigma, \sigma'}\sum_{i_{e,\sigma},j_{h,\sigma'}}  u_{e,\sigma;h,\sigma'}(|{\bf r}_{i_{e\sigma}}-
{\bf r}_{j_{h,\sigma'}}|)\right]\nonumber \\
&=&\exp\left[-\sum_{\sigma}\sum_{i_{e,\sigma},j_{h,\sigma}}  u_{e,\sigma;h,\sigma}(|{\bf r}_{i_{e\sigma}}-{\bf r}_{j_{h,\sigma}}|)-
\sum_{\sigma\ne \sigma'}\sum_{i_{e,\sigma},j_{h,\sigma'}}  u_{e,\sigma;h,\sigma'}(|{\bf r}_{i_{e\sigma}}-{\bf r}_{j_{h,\sigma'}}|)\right]\nonumber 
\\
&\equiv& 
\exp\left[-\sum_{\sigma}\sum_{i_{e,\sigma},j_{h,\sigma}}  \psi(|{\bf r}_{i_{e\sigma}}-{\bf r}_{j_{h,\sigma}}|)-\sum_{\sigma\ne \sigma'}\sum_{i_{e,
\sigma},j_{h,\sigma'}}  \tilde{\psi}(|{\bf r}_{i_{e\sigma}}-{\bf r}_{j_{h,\sigma'}}|)\right].
 \label{Jeh}
 \end{eqnarray}
 \end{linenomath}
\end{adjustwidth}

 Above, we chose one pseudopotential for electron--hole pairs with the same flavor, $u_{e,\sigma;h,\sigma}(r)=\psi(r)$, and another
 for electron--hole pairs with different flavor, $u_{e,\sigma;h,\sigma'}(r)= \tilde{\psi}(r)$. The possibility of $\tilde{\psi}(r)$ being different from $\psi(r)$ should allow the fragmentation of the polyexcitons in smaller components, namely, excitons \cite{Tan, Lee}. 
 This choice, however, breaks the symmetry under an electron--electron pair exchange, as is clear by inspecting Equation~(\ref{Jeh}). Thus, we restore the symmetry by taking the symmetrized  version of $J_{eh}({\bf R}_e,{\bf R}_h)$
\begin{linenomath}
\begin{eqnarray}
J_{eh}^S({\bf R}_e,{\bf R}_h)=
\sum_{P_e} \hat{P}_e \exp\left[-\sum_{\sigma}\sum_{i_{e,\sigma},j_{h,\sigma}}  \psi(|{\bf r}_{i_{e\sigma}}-{\bf r}_{j_{h,\sigma}}|)-
\sum_{\sigma\ne \sigma'}\sum_{i_{e,\sigma},j_{h,\sigma'}}  \tilde{\psi}(|{\bf r}_{i_{e\sigma}}-{\bf r}_{j_{h,\sigma'}}|)\right]
 \end{eqnarray}
 \end{linenomath}
 where $\hat{P}_e$ permutes the electron coordinates. 
 Finally, the trial wave function 
\begin{linenomath}
   \begin{equation}
 \Psi^T_{NX}({\bf R}_e,{\bf R}_h) = J_{ee}({\bf R}_e)J_{hh}({\bf R}_h)J_{eh}^S({\bf R}_e,{\bf R}_h)
\label{eq:guiding_wf}
  \end{equation}
 \end{linenomath}
is nodeless and satisfies all the required symmetry properties: the symmetry under the pair exchange of electrons (holes) and the symmetry under 
the exchange of electrons and holes. We note that in the unsymmetrized electron--hole Jastrow of Equation~(\ref{Jeh}), the pseudopotential $\psi(r)$ describes intraexciton correlations, while $\tilde{\psi}(r)$ describes interexciton correlations.
 
 We take the pseudopotentials of the Pad\'e form. For the electron--electron (hole--hole) pair:
 \begin{linenomath}
   \begin{equation}
  \phi(r)=\frac{c_1 r}{1+c_2 r}.
  \end{equation}
 \end{linenomath}
 
 For the electron--hole pair:
 \begin{linenomath}
\begin{eqnarray}
\psi(r)=\frac{c_3 r+c_4 r^2}{1+c_5 r},\\ 
\tilde{\psi}(r)=\frac{c_3 r+c_6 r^2}{1+c_7 r}.
\end{eqnarray}
 \end{linenomath}
 
Some of the parameters $c_1$--$c_7$ are fixed by exact conditions; the others are determined through an energy minimization as follows: The parameter $c_1$ is fixed by the electron--electron (hole--hole) Kato cusp conditions \cite{Kato}. The parameter $c_3$ is fixed by the electron--hole cusp condition at $d=0$, while for $d>0$, there is no electron--hole cusp and so we set $c_3=0$ \cite{Tan}. We require $c_2,c_5,c_7>0$ to avoid divergences and $c_4,c_6< 0$ to describe the wave function decay when the electron and the hole are far apart. The above wave function should describe separated excitons when either $c_4$ or $c_6$ goes to zero. 

\subsection{Exciton Wave Function} 
The Schr\"odinger equation for the ground state of the isolated exciton, in the center-of-mass reference frame, reads 
\begin{linenomath}
\begin{equation}
\left[ -\frac{2}{r} \frac{\partial}{\partial r} \left( 
r \frac{\partial }{\partial r}\right)+ \frac{2}{\sqrt{r^2+d^2}}\right] \Phi^0_{X}(r)= E^0_X \Phi^0_{X}(r),
\label{eq:X}
\end{equation}  
\end{linenomath}
with $r=|{\bf r}_e-{\bf r}_h|$ the in-plane distance between the electron and the hole. At $d=0$, Equation~(\ref{eq:X}) has the simple hydrogenic normalized solution $\Phi^0_{X}(r)=\sqrt{2/\pi}\exp{[-r]}$, with energy $E^0_X(Ry^*)=-2$. 
However, at $d>0$, there is no closed form for the solution, which must be computed numerically. To this end, we used the simple and 
accurate Numerov algorithm \cite{MS,numerov}, which is especially suited for atomic-like problems.

\section{Results}

In Table \ref{tab:energies}, we report the ground state energy per particle of the exciton, biexciton and quadriexciton as a function of the interlayer distance $d$. The energies of the biexciton and quadriexciton were obtained by DMC simulations using as guiding function the optimized trial function of Equation~(\ref{eq:guiding_wf}). The optimization with respect to the free parameters was performed by minimizing the variational energy using the linear method \cite{linearmethod,linearmethod1} and checking, in selected cases, that the obtained minimum agreed with the one found by the improved stochastic reconfiguration method \cite{sorella-sr}. DMC simulations are affected by the walker population's bias (finite number of walkers $N_w$) and finite time-step bias. However, for small systems, such as those studied here, these biases can easily be made negligible. To this end we used a large number of walkers, $N_w=1760$, and performed an extrapolation to the zero time step. A good optimization and bias reduction are especially important when estimating pair correlation functions.

\begin{table}[h] 
\caption{ Energy per particle $E_X/2 $ (exciton), $E_{2X}/4 $ (biexciton) and $E_{4X}/8$ (quadriexciton) for various distances $d$. Excitonic energies were estimated using the Numerov algorithm \cite{MS,numerov}. The energies for the excitonic complexes were obtained from DMC simulations
with $N_w=1760$ walkers and the time-step bias removed by an extrapolation to the zero time step. 
\label{tab:energies}}
	\begin{tabularx}{\textwidth}{CCCC}
		\toprule
		\boldmath{$d $}	& \boldmath{$E_{X}/2$}	& \boldmath{$E_{2X}/4$}	& \boldmath{$E_{4X}/8$} \\
		\midrule
  	        0.000	& $-$1.00000000	& $-$1.096435(6)	& $-$1.32793(5)	 \\	
                0.100	& $-$0.76594269	& $-$0.80450(4)	& $-$0.89400(5)   \\
        	0.200	& $-$0.64517991 	& $-$0.66372(1)	& $-$0.70988(3)   \\
          	0.300	& $-$0.56511027 	& $-$0.57420(1)	& $-$0.59997(3)   \\
            	0.400	& $-$0.50651906	& $-$0.510687(6)	& $-$0.52533(2)  \\
              	0.500	& $-$0.46112073	& $-$0.46273(1)	& $-$0.47050(1)  \\
                0.550	& $-$0.44192983  	& $-$0.442815(6)	& $-$0.448179(6)   \\
                0.600	& $-$0.42457683   & $-$0.425001(5)	& $-$0.42844(1)   \\
                0.650	& $-$0.40878665 	& $-$0.408919(7)	& $-$0.410877(6)  \\
                0.675   & $-$0.40140710   & $-$0.401466(8)  & $-$0.402841(9)   \\                                              
                0.700	& $-$0.39433924	& $-$0.3943676(9) & $-$0.39498(8)   \\
                0.710	& $-$0.39159457	& $-$0.391622(8)	& $-$0.39202(2)   \\
                0.720	& $-$0.38889512	& $-$0.3889012(5)	& $-$0.38909(1)    \\
                0.740	& $-$0.38362702 	& $-$0.3836290(4)	& -   \\
                0.750	& $-$0.38105610	& $-$0.3810579(5)	& -  \\
                0.760	& $-$0.37852580	& $-$0.3785260(1)	& -   \\
                0.780	& $-$0.37358293	& $-$0.37358300(2) & -   \\
		\bottomrule
	\end{tabularx}
\end{table}

From Table \ref{tab:energies}, it is evident that for $d=0$, we have  $E_X/2> E_{2X}/4 >E_{4X}/8$ and that with an increasing $d$, the energies 
of all three ``phases'' increase, become closer and could presumably cross or merge. So, studying, for instance, the stability of the 
biexciton with respect to the separation into two excitons requires the extrapolation of the available biexciton energies at larger distances $d$. 
In fact, a more efficient manner to study the stability of polyexcitons is to define a binding energy with respect to the separation into fragments. 
The binding energy of a polyexciton is defined as the energy released upon the formation of a bound state with respect to the initially isolated 
constituents (fragments). Therefore, for a biexciton, we can define 
\begin{linenomath}
\begin{eqnarray}
E_B(2X)=2 E_{X}-E_{2X}
\label{eq:2X}
\end{eqnarray}
\end{linenomath}
and on the same footing, the quadriexciton binding energy with respect to two isolated biexcitons is defined as:
\begin{linenomath}
\begin{eqnarray}
E_B(4X)=2 E_{2X}-E_{4X}. 
\label{eq:4X}
\end{eqnarray}
\end{linenomath}

In all the figures with DMC energies, the statistical error are reported, though they are not always visible.

 \subsection{Biexciton Binding Energy}
 Meyertholen and Fogler \cite{Meyertholen_Fogler} showed that when increasing the interlayer distance, the binding energy of a biexciton with respect to two excitons vanished at a finite critical distance $d_c$ and for $d\lesssim d_c$, it had the behavior
 \begin{linenomath}
\begin{eqnarray}
E_B(2X,d)=E_0(2X,d) e^{-D/(d_c-d)},
\label{eq:fogler}
\end{eqnarray}
\end{linenomath}
with 
\begin{equation}
E_0(2X,d)=e^{-6\gamma}/d^4,
\label{eq-E0}
\end{equation}
where $D$ is a positive constant of order one, $\gamma=0.577...$ is the Euler--Mascheroni constant and $D/(d_c-d)\gg 1$. Clearly the last condition implies $E_B\ll E_0$. Their derivation is built on the fact that at at large $d$, the biexciton is weakly bound and the exciton--exciton interaction brings into play at a large interexciton distance the dipole--dipole interaction, present for $d>0$. Clearly the larger the $d$, the larger the exciton dipole and the greater the importance of such an interaction.
 In brief, they neglected what happens at a small interexcitonic distance and wrote a Schr\"odinger equation for two interacting excitons valid at an intermediate distance (first region) and at a large distance (second region): in the first region the binding energy was neglected with respect to the dipole--dipole interaction and in the second, the dipole--dipole interaction was neglected with respect to the binding energy. Imposing the continuity of the logarithmic derivative of the wave function when going from the first to the second region, they obtained an expression involving the binding energy of the biexciton and a characteristic energy of the problem, $E_0(d)$. Further manipulations finally yielded Equation~(\ref{eq:fogler}).
On the basis of their analysis, they proposed a fit of the binding energy in the form
\begin{linenomath}
\begin{eqnarray}
\frac{1}{\log (E_0(2X,d)/E_B(2X,d))}=\frac{d_c-d}{D}+\frac{(d_c-d)^2}{D_1},    
\label{eq:fit} 
\end{eqnarray}
\end{linenomath}
with $d_c,D,D_1$ fitting parameters.

\begin{figure}[H] 
\includegraphics[width=1\textwidth]{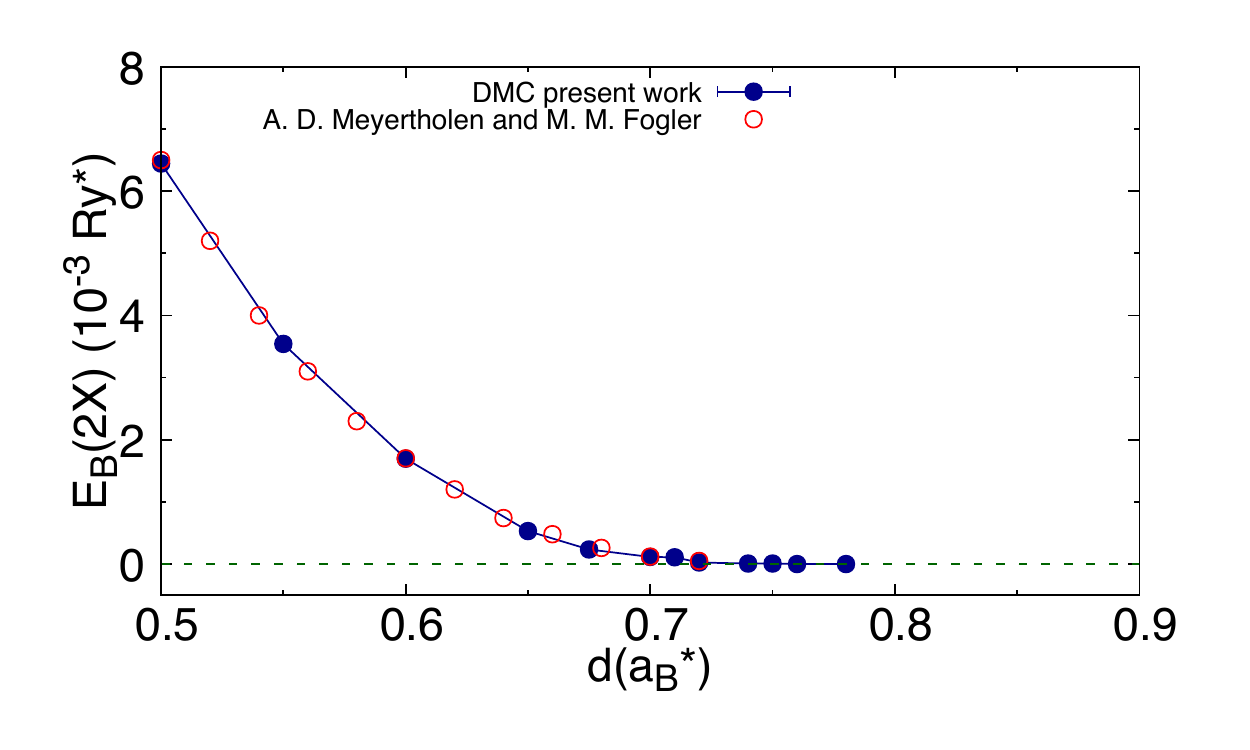}
\caption{ Binding energy  
of the biexciton  $E_B(2X,d)$ (see eq. \ref{eq:2X}) as a function of the interlayer separation $d$.
Together with our results from DMC (solid blue dots), we also report the binding energies from Ref.~\cite{Meyertholen_Fogler} (open red dots). 
The line joining the DMC data is only a guide to the eye.\label{fig:eb-biexc}}
\end{figure}

\begin{figure}[H]
\includegraphics[width=\textwidth]{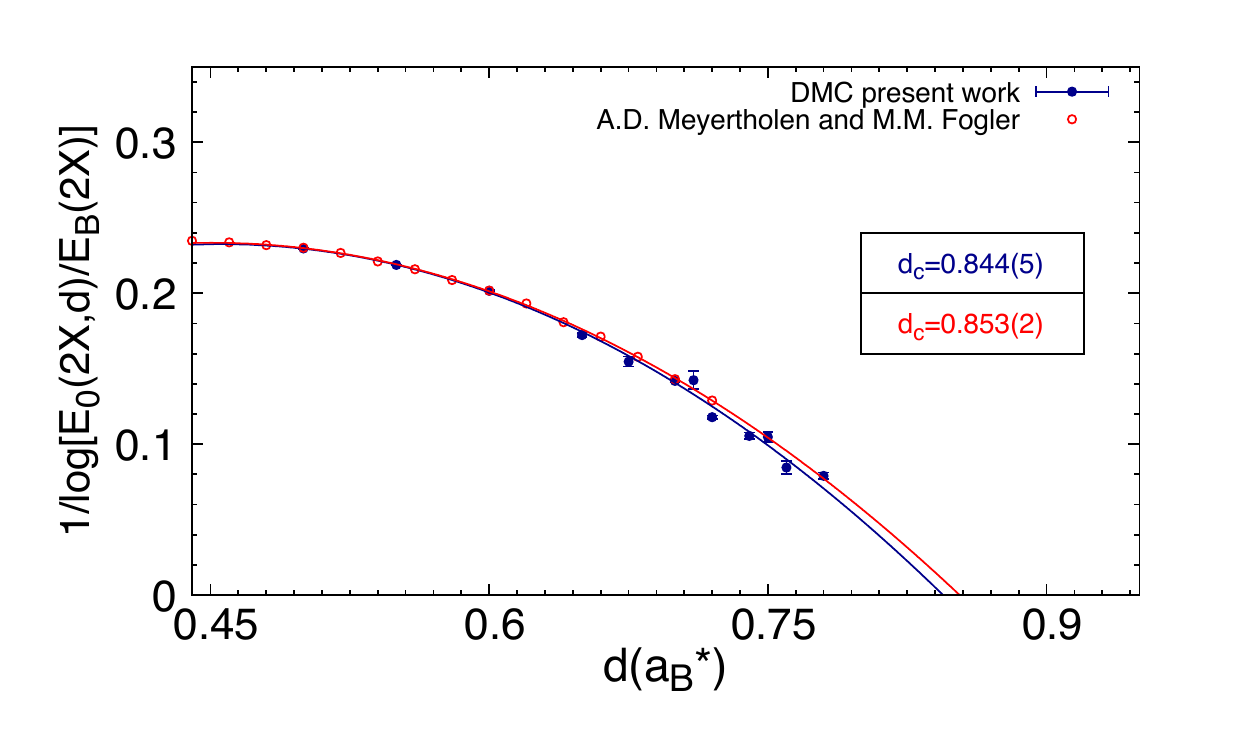}
\caption{  
Logarithmic plot of the biexciton binding energy 
 $E_B(2X,d)$ as a function of the interlayer separation $d$. Data are fitted to the function $1/\log[E_0(2X,d)/E_B(2X,d)]= (d_c-d)/D + (d_c-d)^2/D_1$ as suggested in Ref.~\cite{Meyertholen_Fogler}. We compare results from our DMC simulation and results from Ref.~\cite{Meyertholen_Fogler}. \label{fig:fit-bie}}
\end{figure}

In Figure~\ref{fig:eb-biexc}, we report our DMC results for the binding energy $E_B(2X,d)$ of the biexciton, together with the results of the 
stochastic variational method (SVM)of \cite{Meyertholen_Fogler}. In Figure~\ref{fig:fit-bie}, we report the same data of Figure~\ref{fig:eb-biexc} in a form suitable for the fit according to Equation~(\ref{eq:fit}). One effect of using the ``lens'' provided by the logarithm is to emphasize the region of distances $d$ near $d_c$ as well as the need of extrapolating to $d_c$. It is also clear from Figure~\ref{fig:fit-bie} that the SVM and DMC  results are in good agreement, with the DMC results covering a larger range, which should be important for the fit. Minor differences are found in the values of $d_c$ provided by the fits using the two data sets. We should also stress that in the range of distances covered in Figures~\ref{fig:eb-biexc} and \ref{fig:fit-bie}, $E_0(d) $ is at least two orders of magnitude larger than $E_B$, so that the requirement $E_B\ll E_0$ is fully fulfilled. 
\vspace{-6pt} 

\subsection{Quadriexciton Binding Energy}

\begin{figure}[H] 
\includegraphics[width=\textwidth]{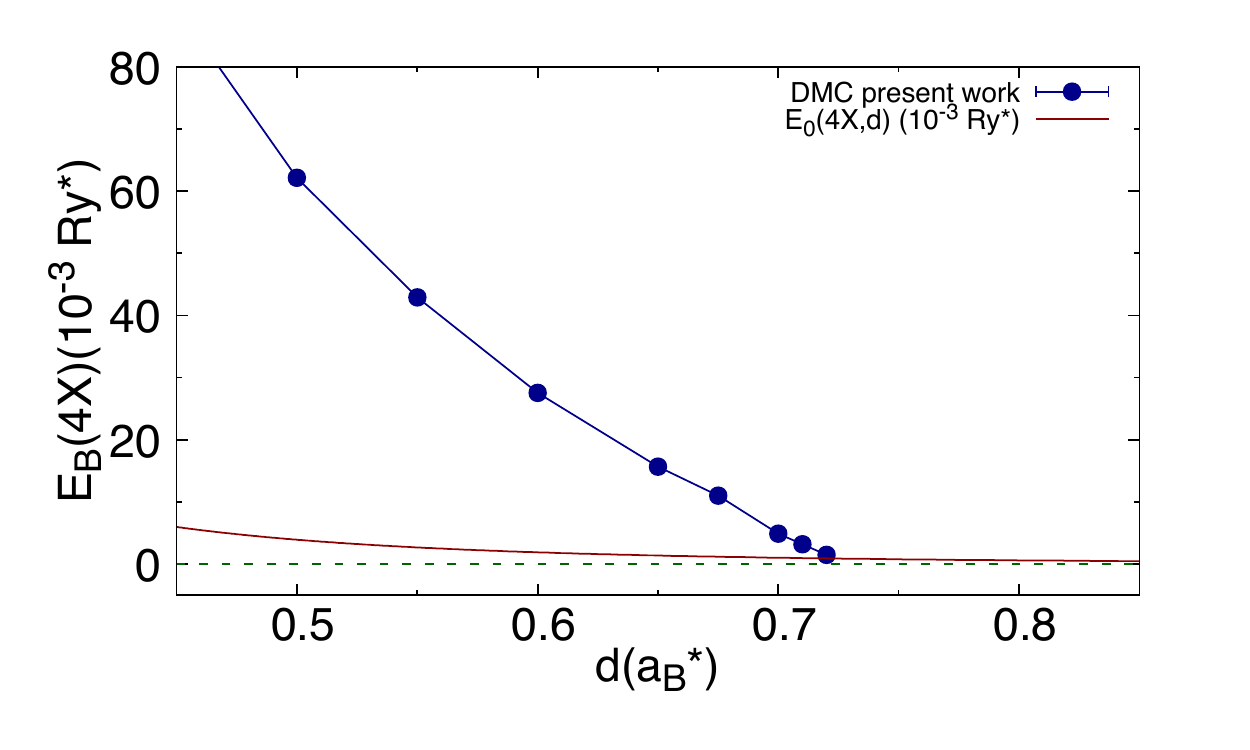}
\caption{ Quadriexciton binding 
energy $E_B(4X,d)$  (see eq. \ref{eq:4X}) as a function of the interlayer separation $d$.
We compare the DMC results (solid blue dots) with $E_0(4X,d)$  (solid dark red line, see eq.  \ref{eq:E0-4X}). The line joining the DMC data is only a guide to the eye.\label{fig:be-qua}}
\end{figure}

\begin{figure}[H]
\includegraphics[width=\textwidth]{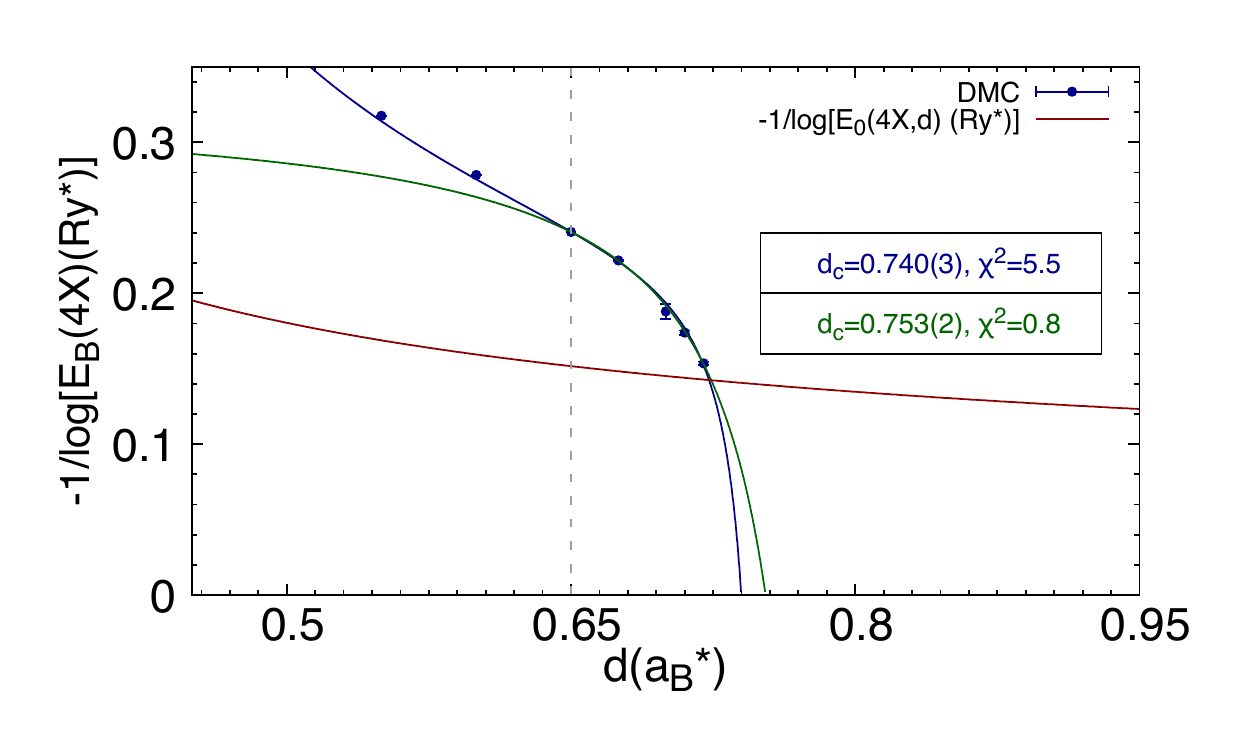}
\caption{
Logarithmic plot of the quadriexciton binding energy 
 $E_B(4X,d)$ as a function of interlayer separation $d$.  $-1/log(E_B(4X,d))$ (solid blue dots) is compared with $-1/log[E_0(4X,d)]$ (solid dark-red line, see eq. \ref{eq:E0-4X}). The DMC data in the range from $d=0.65$ have been fitted to the functions $f(d)=(A/d^4) e^{-D/(d_c-d)}$ (solid blue line) and $q(d)=A e^{-b/(d_c-d)}$ (solid dark green line).
In the box, we report the values of the critical distance for unbinding according to the two fitting functions, together with the reduced $\chi^2$.\label{fig:be-qua1}}
\end{figure}

 In principle, one may study the quadriexciton stability following the treatment for the biexciton \cite{Meyertholen_Fogler}, summarized in some detail in the previous section, also assuming that for the quadriexciton, the binding energy vanishes at a finite critical distance $d_c$. Therefore, at distances $d\lesssim d_c$, the quadriexciton is weakly bound and the biexciton--biexciton interaction (dipole--dipole interaction), comes into play. The Schr\"odinger equation for the relative motion of the two interacting biexcitons is similar to the one for the two interacting excitons, with the following differences. With respect to the exciton--exciton case, the reduced mass is twice larger and the dipole--dipole coupling is four times larger. This yields, when imposing the continuity of the logarithmic derivative of the wave function at large distances (see the biexciton case above), an expression involving the ratio of the binding energy of the quadriexciton $E_B(4X,d)$ and a characteristic energy of the problem which turns out to be 
\begin{equation}
E_0(4X,d)=E_0(2X,d)/128
\label{eq:E0-4X}
\end{equation}
 with $E_0(d)$ given in Equation~(\ref{eq-E0}). However, the self-consistency of the procedure would require $E_B(4X,d)\ll E_0(4X,d)$, a condition which is not fulfilled in the present case, as is clear from Figure~\ref{fig:be-qua}, where our DMC results for $E_B(4X,d)
$ are displayed together with $ E_0(4X,d)$. We therefore fitted functions different from the one used for the biexciton and given in Equation~(\ref{eq:fit}). 
 The first choice was
\begin{linenomath}
\begin{eqnarray}
E_B(4X,d)= A e^{-D/(d_c-d)};
\label{eq:quad-1}
\end{eqnarray}
the second, more flexible, substituted $A$ with the function $A/d^4$; in both cases $A$, $D$ and $d_c$ were the fitting parameters.
\end{linenomath}
\begin{linenomath}
\begin{eqnarray}
E_B(4X,d)= (A/d^4) e^{-D/(d_c-d)};
\label{eq:quad-2}
\end{eqnarray}
\end{linenomath}
\vspace{-18pt}
 The results of the fits to our DMC energies are displayed in Figure~\ref{fig:be-qua1} in terms of the logarithm of $E_B(4X,d)$, to exploit the 
 ``lens'' effect. The two different recipes (the second somewhat inspired by the biexciton fit function) gave compatible results as far as the 
 $d_c$ value was concerned. However, the second choice reproduced the DMC energies in a larger interval of distances, i.e., including also energies that were not included in the fitting.

\subsection{Quadriexciton Pair Correlation Functions}

In a finite system, pair correlation functions can be defined as appropriate ``center-of-mass'' averages of the two-body density. Let us consider 
electron--electron correlations. As we have four electrons (one per flavor), we can start from the two-body electron--electron~density
 \begin{linenomath}
\begin{eqnarray}
\rho^{\sigma\sigma'}_{ee}({\bf r}_1,{\bf r}_2)=\langle \delta ({\bf r}_1- {\bf r}_{e,\sigma}) \delta ({\bf r}_2- {\bf r}_{e,\sigma'})\rangle,
\label{eq:rho-1}
\end{eqnarray}
\end{linenomath}
where $\langle ... \rangle$ indicates a ground state average and only the different-flavor case is present. We define new vectors as ${\bf R}
=({\bf r}_1+{\bf r}_2)/2$ and ${\bf r}={\bf r}_1-{\bf r}_2$, so that ${\bf r}_1={\bf R}+{\bf r}/2$ and ${\bf r}_2={\bf R}-{\bf r}/2$. Substituting in Equation~(\ref{eq:rho-1}), we get 
\begin{linenomath}
\begin{equation}
\rho^{\sigma\sigma'}_{ee}({\bf R}+{\bf r}/2,{\bf R}-{\bf r}/2)=\langle \delta ({\bf R}+{\bf r}/2- {\bf r}_{e,\sigma}) \delta ({\bf R}-{\bf r}/2- {\bf r}_{e,
\sigma'})
\rangle,
\label{eq:rho-2}
\end{equation}
\end{linenomath}
and integrating over the ``center-of-mass'' vector ${\bf R}$, we obtain a pair correlation function 
\begin{linenomath}
\begin{equation}
g^{\sigma\sigma'}_{ee}({\bf r} )=\langle \delta ({\bf r}- {\bf r}_{e,\sigma}+{\bf r}_{e,\sigma'}) \rangle.
\label{eq:g-2}
\end{equation}
\end{linenomath}

A few observations are in order. First, $g^{\sigma\sigma'}_{ee}({\bf r} )$ is a dimensional quantity; it has the dimensions of a density. However, for our finite system, there is no sensible density to divide by to get a dimensionless pair correlation function, as in extended systems. The integration  of $g^{\sigma\sigma'}_{ee}({\bf r} )$ with respect to ${\bf r}$ over the whole space is one. One usually finds convenient to take the angles' average of $g^{\sigma\sigma'}_{ee}({\bf r} )$ to get a $g^{\sigma\sigma'}_{ee}({ r} )$. In a similar manner, one may define electron--hole pair correlation functions as
\begin{linenomath}
\begin{equation}
g^{\sigma\sigma'}_{eh}({\bf r} )=\langle \delta ({\bf r}- {\bf r}_{e,\sigma}+{\bf r}_{h,\sigma'}) \rangle,
\label{eq:geh-1}
\end{equation}
\end{linenomath}
having in this case both same-flavor $\sigma'=\sigma$ and different-flavor $\sigma'\neq\sigma$ functions. We should mention that the above definitions for the pair correlation functions coincide with that of the intracule density in chemistry as defined in \cite{Umrigar2007}.

\begin{figure}[H] 
\includegraphics[width=.8\textwidth]{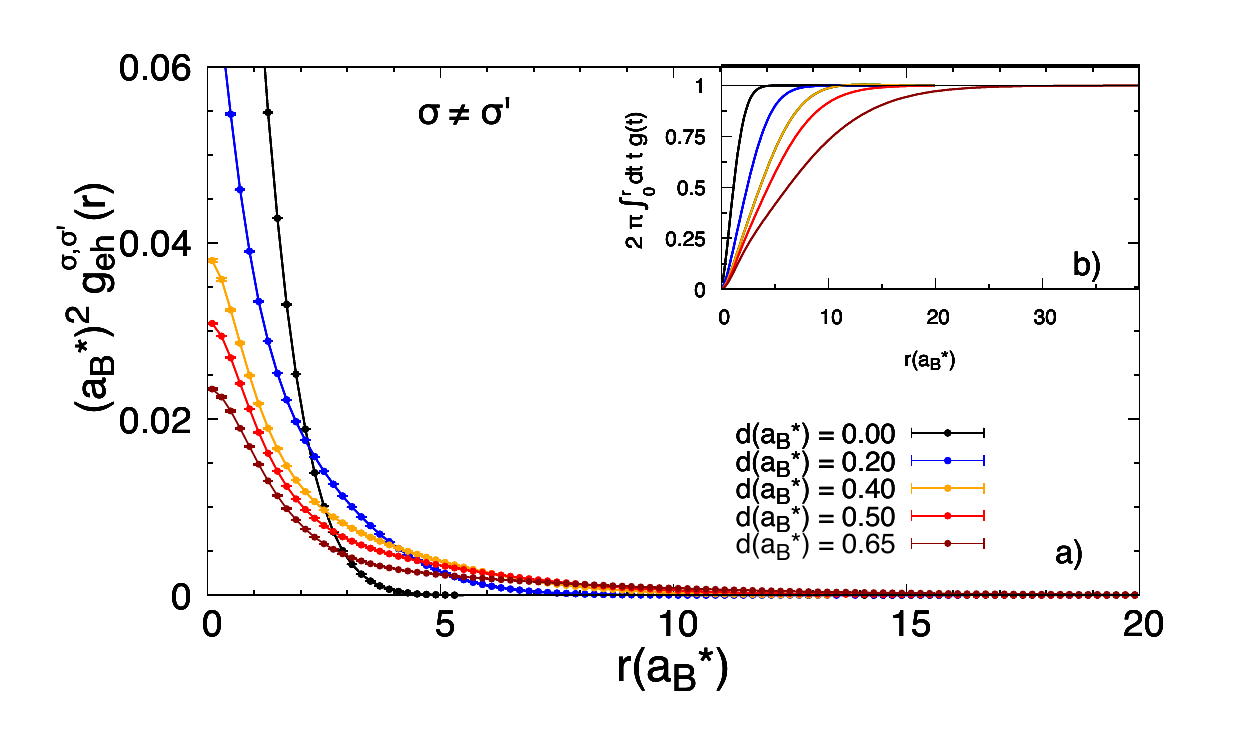}
\caption{ Electron--hole pair correlation 
functions for the quadriexciton at several distances $d$. In panel (\textbf{a}), extrapolated DMC $g_{eh}(r)$ values are shown for distances $d=0.0, 0.2, 0.4, 0.5$ and $0.65 (a_B^*)$ with black, blue, orange, red and dark red solid points, respectively.
Lines joining the DMC data are only a guide to the eye.
In panel (\textbf{b}), we show the quantity $2 \pi \int^r_0 dt t g(t)$ that sums up to $1$ in all cases for the $r$-ranges considered here.}
\label{fig:gr_eh_qua}
\end{figure}

\begin{figure}[H] 
\includegraphics[width=.8\textwidth, trim = 0 5 0 0]{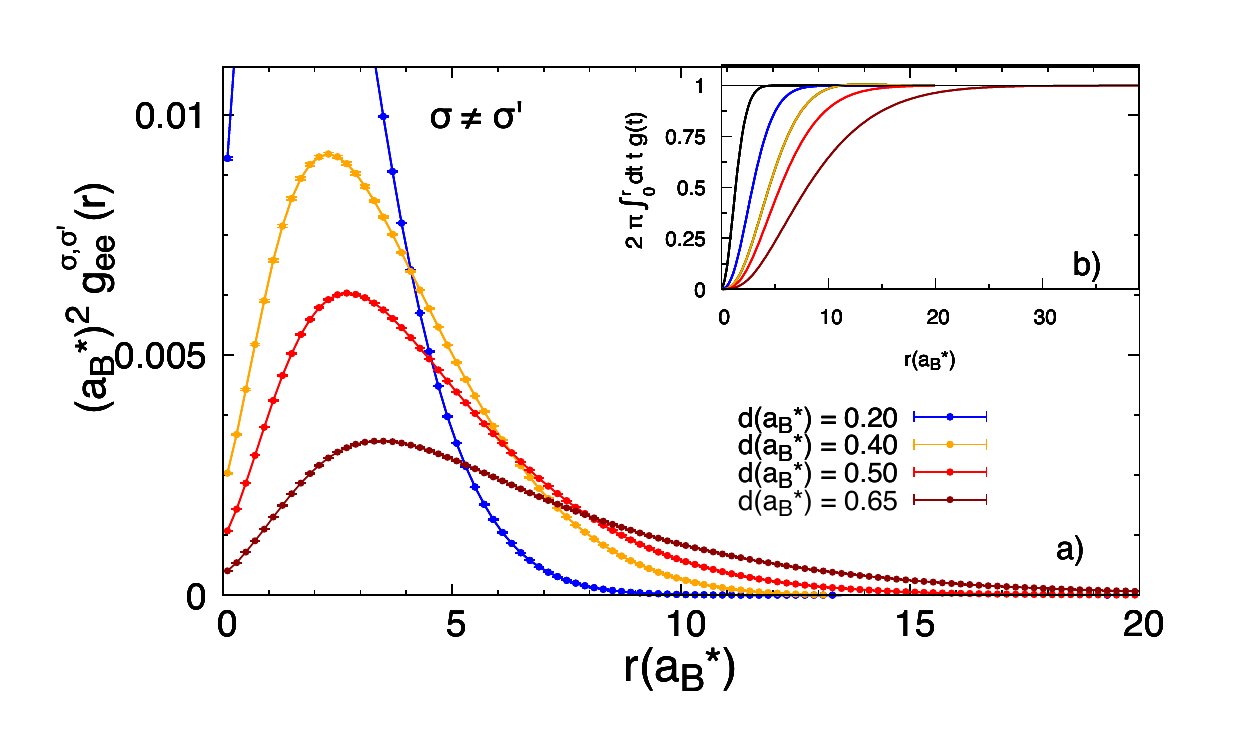}
\caption{ Electron--electron pair correlation
functions for the quadriexciton at several distances $d$. In panel (\textbf{a}), extrapolated DMC $g_{ee}(r)$ values are shown for distances
$d=0.0, 0.2, 0.4, 0.5$ and $0.65 (a_B^*)$ with black, blue, orange red and dark red solid points, respectively.
Lines joining the DMC data are only a guide to the eye. In panel (\textbf{b}), we show the quantity $2 \pi \int^r_0 dt t g(t)$ that sums up to $1$ in all cases for the $r$-ranges considered here.}
\label{fig:gr_ee_qua}
\end{figure}

In Figure~\ref{fig:gr_eh_qua}, we show the electron--hole pair correlation functions, at various distances $d$, averaged on the azimuthal angle. 
Clearly, the contact value $g^{\sigma\sigma'}_{eh}( r=0 )$ is a decreasing function of $d$. At the largest $d$ shown ($d=0.65$), there is still an appreciable pileup of probability at the origin. In Figure~\ref{fig:gr_ee_qua}, we show the electron--electron pair correlation functions. A peak at distances on the scale of a few Bohr radii is present, due to the quadriexciton binding; similarly to what happens to the contact values 
of $g^{\sigma\sigma'}_{eh}( r )$, such a peak is a decreasing function of $d$.  Our results were similar to those in \cite{Lee} for the biexciton.  
One should keep in mind that different units were used in the two studies, so our pair correlation functions should be multiplied by four, before 
comparing with those of the biexciton case in \cite{Lee}. Finally, as a symmetry check, we verified that the electron--hole pair correlation 
functions with the same or different flavor coincided; similarly, the electron--electron pair correlation functions did not depend on the pair of different flavors chosen.
\vspace{-12pt}

\section{Discussion}
We studied a system of four electrons and four holes (quadriexciton) in a symmetric, paramagnetic bilayer, where indirect excitons are formed, 
focusing on (i) the unbinding of the quadriexciton into two biexcitons and (ii) the pair correlation functions. To this end, we used state-of-the-art QMC simulations in which both quadriexcitons and biexcitons were studied. We already analyzed the various properties of biexcitons and quadriexcitons in the foregoing. Here, we simply discuss what are the implications of our results and analysis in relation to the phase diagram of the symmetric, paramagnetic bilayer at a finite density \cite{ maezono2013,quadri}. 

According to our analysis and the one in \cite{Meyertholen_Fogler}, an isolated biexciton unbinds into two excitons at $d_c=0.84a_B^*$. 
Evidently, one might speculate that such result could be of some relevance to the condensed phase, though only at a very small density. The phase 
diagram in \cite{ maezono2013} covers the density range $0\leq r_s \leq 8$ and reveals a biexcitonic phase only for $4 \lesssim r_s\lesssim 
8 $ and $d  \lesssim d_{2X}(r_s) = 0.05a_B^*$. One may therefore conclude that at such densities the electronic screening is very effective 
in destabilizing the biexcitonic phase and favoring its melting into excitons, to the point that the biexcitonic phase almost disappears. 
The situation appears qualitatively different in the symmetric, paramagnetic bilayer with valley degeneracy \cite{quadri}. In this system, a 
robust quadriexcitonic phase is present in the same density range $0\leq r_s \leq 8$. It first appears around $r_s=1.5$ at $d=0$ but with 
increasing $r_s$, the quadriexciton--exciton boundary increases up to $d\simeq 0.65 a_B^*$ at $r_s=8$. This $d$ value is just 12\% smaller 
than the $d=0.74 a_B^*$ at which the isolated quadriexciton melts into two biexcitons. One may then observe, comparing Figures~\ref{fig:eb-biexc} and \ref{fig:be-qua}, that the binding energy of the quadriexciton is about one order of magnitude larger than that of the biexciton for $0.5a_B^* \leq d \leq 0.65 a_B^*$. Is that the reason of the much greater stability of the quadriexciton in the condensed phase, as compared with the biexciton? We plan to further study the properties of polyexcitons in the near future to answer this and other questions.

Our goal in this paper was to provide benchmark simulation results for the simplest model electron--hole bilayer when valley degeneracy was added. Though we were not trying to accurately model real devices, we could nevertheless link our results to fabricated coupled graphene bilayers, by estimating the effective Bohr radius $a_B^*$ in actual devices and hence $d_{exp} / a_B^*$, where $d_{exp}$ is the actual inter-bilayer distance. This procedure relied on crude assumptions on the form of the interparticle interactions, as well as on the chosen values of the dielectric constants and band masses. We gave three examples, using the isotropic interactions in Equation~(\ref{H}). In \cite{tutucprl2018}, coupled graphene bilayers with $d_{exp}=2.2$ nm were studied. Assuming $m_b=0.04m_e$ and $\epsilon=3$ or $\epsilon=14$ (the smallest or the largest of the dielectric constants of the constituent materials), we obtained $d_{exp} / a_B^*=0.55$ or $d_{exp} / a_B^*=0.12$. In  \cite{deanGBprl}, coupled graphene bilayers with 5 nm$\leq d_{exp} \leq 12$ nm were investigated. Choosing $m_b=0.04m_e$ and $\epsilon=3$, the authors obtained $1.3\leq d_{exp}/a_B^* \leq 3.0$. 
Finally, in \cite{tutucGBprl}, where  devices with the same band mass and dielectric constant as in \cite{deanGBprl} were studied, with $2nm\leq d_{exp} \leq 5nm$, the authors obtained $0.50\leq d_{exp}/a_B^* \leq 1.3$. Comparing the estimates of $d_{exc}/a_B^*$ for the three sets of experiments \cite{tutucprl2018,deanGBprl,tutucGBprl} with the results for our model electron--hole bilayer, one would conclude that only the devices studied in \cite{tutucprl2018,tutucGBprl}  would be compatible with the observation of an isolated quadriexciton.




\vspace{6pt} 



\authorcontributions{Conceptualization, G.S. and S.D.P.;  methodology, G.S. and S.D.P.; investigation, S.D.P.  and C.M. 
All the authors discussed the details of the work and contributed to the writing of the paper.
 All authors have read and agreed to the published version of the manuscript.}

\funding{This research received no external funding.}
\institutionalreview{
}

\informedconsent{

}
\dataavailability{Data are available on request.} 


\conflictsofinterest{The authors declare no conflict of interest.} 
\clearpage 

\begin{adjustwidth}{-\extralength}{0cm}
\reftitle{References}

\PublishersNote{}
\end{adjustwidth}
\end{document}